\documentclass{article}

\usepackage{import}

\usepackage{algorithm}
\usepackage{algpseudocode}

\usepackage{listings}
\usepackage{xcolor}
\usepackage{hyperref}
\usepackage{subcaption}
\usepackage{graphicx}

\usepackage{soul,color}
\usepackage[multiple]{footmisc}
\usepackage{xurl}
\usepackage{mathtools}
\usepackage{caption}

\usepackage[a4paper]{geometry}
\newgeometry{left=1.2in,bottom=1.2in,right=1.2in,top=1.2in}

\definecolor{codegreen}{rgb}{0,0.6,0}
\definecolor{codegray}{rgb}{0.5,0.5,0.5}
\definecolor{codepurple}{rgb}{0.58,0,0.82}
\definecolor{backcolour}{rgb}{0.95,0.95,0.92}

\lstdefinestyle{mystyle}{
  backgroundcolor=\color{backcolour},   commentstyle=\color{codegreen},
  keywordstyle=\color{magenta},
  numberstyle=\tiny\color{codegray},
  stringstyle=\color{codepurple},
  basicstyle=\ttfamily\footnotesize,
  breakatwhitespace=false,         
  breaklines=true,                 
  captionpos=b,                    
  keepspaces=true,                 
  numbers=left,                    
  numbersep=5pt,                  
  showspaces=false,                
  showstringspaces=false,
  showtabs=false,                  
  tabsize=2
}

\AtBeginDocument{%
  \providecommand\BibTeX{{%
    \normalfont B\kern-0.5em{\scshape i\kern-0.25em b}\kern-0.8em\TeX}}}

\begin{document}

\title{Tag Your Fish in the Broken Net: A Responsible Web Framework for Protecting Online Privacy and Copyright}

\author{Dawen Zhang$^{1,2}$, Boming Xia$^{1}$, Yue Liu$^{1}$, Xiwei Xu$^{1}$,\\
Thong Hoang$^{1}$, Zhenchang Xing$^{1,2}$, Mark Staples$^{1}$,\\
Qinghua Lu$^{1}$, Liming Zhu$^{1}$
\\
\\
\textit{CSIRO's Data61$^1$}
\\
\textit{Australian National University$^2$}
\\
\texttt{David.Zhang@data61.csiro.au}
}
\date{}
\maketitle

\begin{abstract}
The World Wide Web, a ubiquitous source of information, serves as a primary resource for countless individuals, amassing a vast amount of data from global internet users. However, this online data, when scraped, indexed, and utilized for activities like web crawling, search engine indexing, and, notably, AI model training, often diverges from the original intent of its contributors. The ascent of Generative AI has accentuated concerns surrounding data privacy and copyright infringement. Regrettably, the web's current framework falls short in facilitating pivotal actions like consent withdrawal or data copyright claims. While some companies offer voluntary measures, such as crawler access restrictions, these often remain inaccessible to individual users. To empower online users to exercise their rights and enable companies to adhere to regulations, this paper introduces a user-controlled consent tagging framework for online data. It leverages the extensibility of HTTP and HTML in conjunction with the decentralized nature of distributed ledger technology. With this framework, users have the ability to tag their online data at the time of transmission, and subsequently, they can track and request the withdrawal of consent for their data from the data holders. A proof-of-concept system is implemented, demonstrating the feasibility of the framework. This work holds significant potential for contributing to the reinforcement of user consent, privacy, and copyright on the modern internet and lays the groundwork for future insights into creating a more responsible and user-centric web ecosystem.
\\
\\
Keywords: user consent, privacy, copyright, web crawler, transparency, responsible web

\end{abstract}

\section{Introduction}

The World Wide Web (WWW) has revolutionized the way people communicate, learn, and share knowledge. From its inception as a tool for academic collaboration~\cite{berners1999weaving} to its evolution into a global platform intertwining with %
our daily lives, the WWW has become an indispensable part of modern society. Yet, as it has grown in scale and complexity, so too have the challenges associated with managing and safeguarding the vast amounts of data it contains. While the WWW has democratized information access, it has also given rise to concerns about how this information, which is often contributed by users from all corners of the globe, is repurposed and utilized.
A prominent example of this repurposing is the training of Generative AI (GenAI) models~\cite{khan2022subjects}, such as Large Language Models (LLMs).
These models, while transformative in their capabilities to generate content based on patterns in existing data, bring to the fore pressing concerns about data privacy, copyright infringements, and the broader ethics of data usage~\cite{hacker2023regulating}.

Central to these concerns is the EU's General Data Protection Regulation (GDPR)~\cite{gdpr}, a pivotal regulatory framework safeguarding data protection and sovereignty. Established in 2018, the GDPR outlines the rights of EU citizens concerning their personal data and sets rigorous standards for entities engaged in data collection or processing. However, the dynamic nature of GenAI models, especially in their data sourcing and utilization, poses unique challenges in ensuring privacy compliance. Such concerns have led to various investigations or class actions\footnote{\url{https://www.forbes.com/sites/emmawoollacott/2023/09/01/openai-hit-with-new-lawsuit-over-chatgpt-training-data/}}.
Concurrently, as the digital domain burgeons, copyright concerns have also gained prominence. The unauthorized harnessing of copyrighted content, especially when curating training datasets for GenAI models like LLMs, has precipitated notable legal confrontations. Cases involving GitHub Co-pilot accused of redistributing copyrighted code without proper attribution\footnote{\url{https://www.theregister.com/2023/05/12/github_microsoft_openai_copilot/}}, Stable Diffusion allegedly copied over 12 million images from Getty Images without permission\footnote{\url{https://www.theverge.com/2023/2/6/23587393/ai-art-copyright-lawsuit-getty-images-stable-diffusion}}, and ChatGPT sued by U.S. authors for misusing their writings in training\footnote{\url{https://www.reuters.com/technology/more-writers-sue-openai-copyright-infringement-over-ai-training-2023-09-11/}}.
Regrettably, the current infrastructure of the WWW is ill-equipped to effectively tackle these multifaceted challenges (see detailed discussion in Section~\ref{Sec:Challenges}).

To address these pressing concerns
amidst the burgeoning era of GenAI and LLMs on the WWW, this paper introduces a novel, user-centric consent tagging framework. This framework leverages the extensibility of HTTP and HTML in conjunction with the decentralized nature of distributed ledger technology like Blockchain. Our primary objective is to provide users with enhanced control over their online data, enabling them to tag, monitor, and, if necessary, request data withdrawal from data custodians.

In weaving together these techniques, this research endeavors to harmonize the transformative potential of GenAI with the foundational principles of data privacy and user empowerment. Ultimately, this work contributes to the vision of a web ecosystem that prioritizes responsibility and user autonomy. The primary contributions of this paper are listed as follows:
\begin{itemize}
    \item We underscore the challenges and technical disparities between the present Web infrastructure and the legal mandates for privacy and copyrights, with a specific focus on the data practices of Generative AI models;
    \item We introduced a novel architecture that augments the existing World Wide Web framework, facilitating responsible user-centric data practices including data collection, data scraping, and data distribution;
    \item We demonstrate the feasibility of our framework, highlighting its compatibility, extensibility, and seamless integration with the current Web infrastructure.
\end{itemize}

The remainder of this paper is organized as follows.
Section~\ref{sec:motivation} explains the motivating scenario of this study, while Section~\ref{sec:design} presents the overall design of architecture and system mechanism, with the subsequent section demonstrating the implementation details. Section~\ref{sec:evaluation} evaluates the proposed approach, followed by Section~\ref{sec:discussion} which discusses insights and sheds light on the study's limitations. A review of the existing literature related to this work is presented in Section~\ref{sec:related-work}, and Section~\ref{sec:conclusion} concludes the paper.

\section{Motivating Scenario}
\label{sec:motivation}

The current infrastructure of WWW allows the utilization of online data with nearly no restrictions for companies but fails to provide transparency and control for users. This section demonstrates the motivating scenario by describing a typical journey of user data, followed by discussing the challenges of building a responsible web.

\subsection{Data Journey: From User Activities to Training AI models}

\begin{figure}[htbp]
    \centerline{\includegraphics[width=0.8\columnwidth]
    {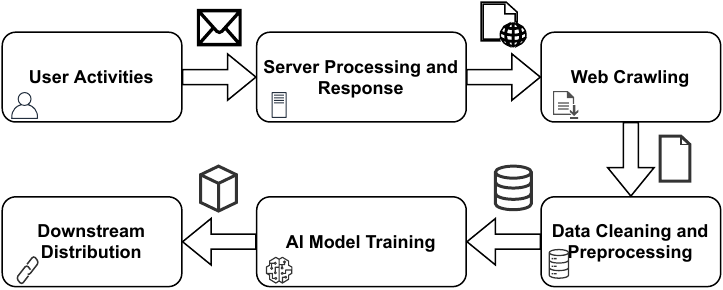}}
    \caption{Illustration of the User Data Journey, from User Activities to Downstream Distribution.}
    \label{data-journey}
\end{figure}

This section describes a typical journey of user data, tracing down its path from initial web-based activities to its utilization in training AI models, as illustrated in Figure \ref{data-journey}.

\noindent \textbf{User Activities.}
Internet users interact with the web through various interfaces, termed as \textit{clients}, including web browsers and mobile applications. These interactions are predominantly facilitated by \textit{HTTP requests}. When users, for instance, upload a selfie via a web browser or a mobile app, the client initiates an HTTP request. This request generally contains the data in its body and is dispatched to the designated \textit{server}. While HTTP encompasses methods for diverse types of requests, data transmission is commonly achieved through methods like \textit{POST}, \textit{PUT}, and \textit{PATCH}.

\noindent \textbf{Server Processing and Response.}
Upon receiving the user-submitted data via HTTP request from a client, the web server processes the request. The server routes the requests to specific \textit{handlers} based on a \textit{routing} mechanism. The routing typically takes into account the HTTP method (e.g. GET, POST, PUT) and parses the URL to direct requests to the correct handlers. The handler further processes the request, which might extract various pieces of information from the request, including the client's user-agent, user-submitted data, and other relevant details from both the request header and body. Depending on the purpose and relevance, the extracted data might be validated and saved in the \textit{databases or file systems}.

When a resource, such as a web page or an \textit{API} endpoint containing pertinent data, is being accessed, data is fetched from storage or cache, embedded in an HTML page or payload (e.g., JSON), and subsequently sent by a server in an HTTP GET \textit{response} back to the client, which then renders or processes it accordingly.

\noindent \textbf{Web Crawling.}
\textit{Web crawlers}, operated by entities such as search engines, researchers, and archiving services, automatically browse and collect web page information. They continuously crawl the web pages of websites by sending HTTP GET requests to fetch the content of URLs. The crawler downloads the web page content, which is in response to its request, and attributes certain information to the downloaded page, such as the URL and metadata~\cite{dikaiakos2005investigation}.

To avoid overloading website servers, the crawlers often use an adaptive back-off algorithm to introduce a delay between consecutive requests to the same server~\cite{manning2008introduction}. Additionally, as documented in Robots Exclusion Protocol (extended by proposed RFC 9309\footnote{\url{https://www.rfc-editor.org/rfc/rfc9309}}), the web crawlers should respect \textit{robots.txt}, which specifies %
which pages or directories should not be crawled. OpenAI also launched \textit{GPTBot}\footnote{\url{https://platform.openai.com/docs/gptbot}}, and provides specific configurations and \textit{IP ranges} to allow website administrators to control its access to their sites. Google recently followed to announce \textit{Google-Extended}\footnote{\url{https://blog.google/technology/ai/an-update-on-web-publisher-controls/}} to give web publishers certain control over their web crawling activities.

\noindent \textbf{Data Cleaning and Preprocessing.}
Once data is collected, especially from the vast expanse of the web, it often requires \textit{cleaning and preprocessing} before being deployed for analytics and AI model training. For example, data might be anonymized to safeguard the privacy of web users~\cite{brickell2008cost}. Furthermore, 
web-sourced data can be messy—it might have missing fields, duplicate records, or inconsistencies~\cite{manku2007detecting}. Cleaning ensures these issues are addressed. After cleaning, the data might still not be ready for analysis or AI training. Preprocessing steps, such as normalizing numerical values, encoding categorical variables, or deriving new insightful fields, transform the data into a more usable format~\cite{garcia2015data}.

\noindent \textbf{AI Model Training.}
With the data cleaned and preprocessed, it can then be used to train AI models. Depending on the training paradigm, preparatory steps such as data labeling might be essential. The choice of model architecture and training techniques hinges on the specific domain and objective~\cite{roh2019survey}. Currently, if data needs to be removed from a trained model, the process often employs retraining or machine unlearning methods~\cite{7163042,5484614}.

\noindent \textbf{Downstream Distributions}
After the model is trained and deployed, there are still potential \textit{downstream distributions} and uses of the knowledge derived from original data~\cite{jaderberg2014synthetic}. While these downstream distributions typically do not involve sharing raw user data directly, %
they can inadvertently reveal information related to the original user data, especially due to the issues such as \textit{model memorization}~\cite{carlini2021extracting} and \textit{membership inference attacks}~\cite{shokri2017membership}. Examples of such downstream distributions include outputs of the model, sharing of pre-trained model weights, and incorporation of model outputs into datasets for training new models.

\subsection{Challenges of Protecting User Privacy and Copyrights}
\label{Sec:Challenges}

Online content often includes personal data subject to regulations like the GDPR~\cite{wachter2019data}. However, integrating this data into (Gen)AI models introduces issues due to their diverse and expansive data sources~\cite{zhang2023right}. For example, While service providers might notify users of data collection, data harvested from the internet for GenAI training often bypasses this notification process, violating the \textit{Right to be Informed (Art.~13, Art.~14)}. After the collection, data is integrated into GenAI models without providing the information about the dataset, making it difficult to access the original data, which may violate the \textit{Right of Access (Art.~15)}. In addition, despite users' \textit{Right to (Withdraw) Consent (Art.~7)} to data processing, the potential ambiguities surrounding data collection and usage by GenAI models might complicate the clear acquisition and withdrawal of user consent. Given the swiftly advancing nature of the technology, the original consent may no longer align with the model's evolving usage and the emerging capabilities of these models.

Similarly, when considering the copyright of online content, particularly that shared by artists on websites, existing internet infrastructure lacks a straightforward mechanism to identify the original content creators for the purposes of seeking consent or offering compensation.

The current architecture of the WWW presents significant obstacles for companies in upholding these rights. Even if companies are genuinely committed to guaranteeing them, the practical implementation proves challenging~\cite{renieris2023beyond}. For instance, when data is crawled from the web, firms cannot notify data subjects about this activity since they lack direct user contact. Additionally, verifying an individual's authenticity without requesting supplementary information becomes untenable, further complicating the process of consent withdrawal.

Due to these challenges as well as the increasing scrutiny from regulatory bodies and persistent advocacy from affected stakeholders, organizations have been prompted to adopt more transparent practices. OpenAI, for instance, published blog posts about their AI safety practice and released documentation about their GPTBot and methods for web administrators to disallow or customize their crawling activities. Though pushed by the public, such voluntary self-regulation demonstrates a commitment to the responsible web and the ethical development of AI. Similarly, Google followed to announce Google-Extended, which provided a similar approach for web administrators to control the access of crawlers.

However, these approaches, which largely rely on the Robots Exclusion Protocol, are insufficient in building a truly responsible web that robustly safeguards the privacy and copyrights of online users. These methods offer only coarse-grained control over crawler access, typically restricting access to directories and files rather than specific content on a page. Additionally, the right to enforce these site-level policies is granted to website administrators rather than the users. This becomes especially problematic on social media or content-sharing platforms where content creators' rights and autonomy are neglected. These limitations underscore the need for a more comprehensive and user-centric approach.

\section{Architectural Design}
\label{sec:design}

To address these challenges and provide a comprehensive and user-centric solution for a responsible web, we introduce a novel framework. We present the architectural design of the proposed framework in this section. The system architecture and its process interactions are described in~\ref{sec:system-arch} and~\ref{sec:system-mechanism}, respectively.

\subsection{System Architecture}
\label{sec:system-arch}

\begin{figure*}[htbp]
\centerline{\includegraphics[width=0.9\textwidth]
{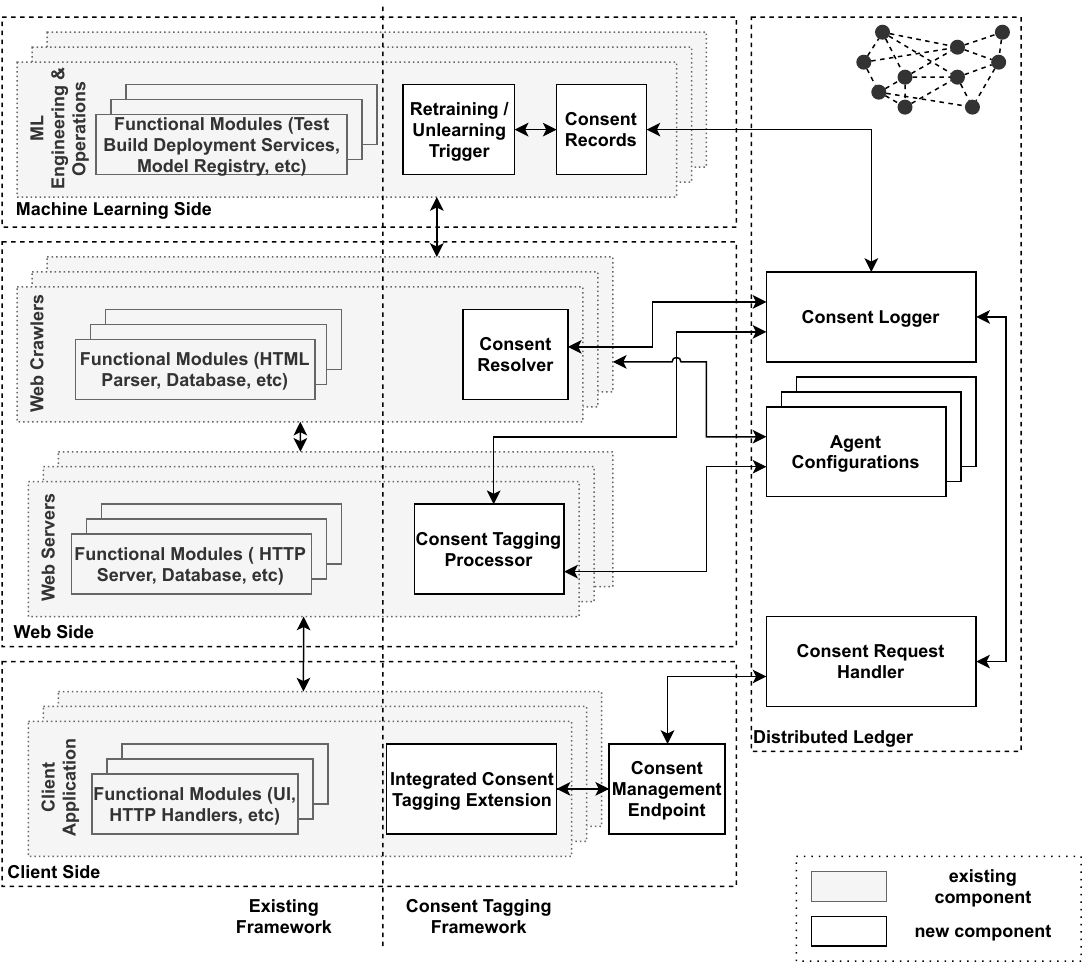}}
\caption{System Architecture. The framework features four layers, including the Client Side, Web Side, Machine Learning Side, and Distributed Ledger.}
\label{architecture}
\end{figure*}

Fig.~\ref{architecture} demonstrates the overview of our proposed system architecture, which consists of four main layers: client, web, machine learning, and distributed ledger sides. Please note that the proposed architecture also contains components in conventional web application architectures, while in this paper we focus on the consent tagging framework for data privacy and user empowerment.

\noindent \textbf{Client Side.}
The client layer contains three main components: the functional modules of client applications, the Integrated Consent Tagging Extension, and a Consent Management Endpoint. The functional modules provide functionalities of the client application, such as user interfaces, network communications, and computations. The Integrated Consent Tagging Extension serves as an extension inside the client application, through the forms of browser extension, application plugin, or software library. Upon any HTTP request being sent, the Integrated Consent Tagging Extension injects a \textit{tag} as an additional HTTP header in the request.
\vspace{-0.5em}
\begin{table}[h!btp]
\centering
\footnotesize
\begin{tabular}{|l|}
\hline
Request URL: https://www.reddit.com/submit\\
Request Method: POST\\
\hline
\hline
HTTP Request Headers\\
\hline
Content-Type: application/x-www-form-urlencoded\\
User-Agent: Mozilla/5.0 (Windows NT...\\
X-Consent-Config: GPTBot:0;Googlebot:1;default:0\\
X-Consent-Tag-Hash: ab4a39e4fc8118cbb37c...\\
X-Consent-Tag-Sig: 646d3d1079b5ac19ea5b...\\
...\\
\hline
\end{tabular}
\caption{Example HTTP request headers of user post. The consent configuration, the hash, and the signature of the data are included in the headers.}
\label{table:httpheader}
\end{table}
\vspace{-0.75em}

The header contains two parts: a consent configuration of which crawlers are consented, and a tag containing two parts, which further include the hash $H_d$ of the user data, and the digital signature $S_d$ of the hash $H_d$ using a key pair $K$. An example of the headers is shown in Table \ref{table:httpheader}. The metadata of the request, the hash $H_d$, and the signature $S_d$ are stored locally. Optionally, the client-side program that initiated the post request can specify whether the request contains data that will be stored on the server using a header, which the Integrated Consent Tagging Extension can check upon the request being sent, to decide whether tag and log this request or not.

A user can track the data through the Consent Management Endpoint. The endpoint will query the distributed ledger to extract the journey of the consent tag. If the user wishes to withdraw the consent, the endpoint can broadcast the consent withdrawal request on the distributed ledger. In order to prove the ownership of the consent tag, the withdrawal request contains the hash $H_d$, digital signature $S_d$, public key $K_{pub}$ of key pair $K$, and a signature $S_c$ signed by private key $K_{pri}$ of key pair $K$ on a challenge from the random number generator on the distributed ledger.

\noindent \textbf{Web Side.}
After receiving the HTTP requests from the client, the web server will proceed with a series of operations to fulfill its functionalities. The consent tags should be linked with the corresponding data stored in the database. Once the data is provided through APIs or HTML pages, the Consent Tagging Processor will embed the tag in the data. The tag is either provided in a designated field in the document such as JSON, or as an HTML DOM attribute. An example HTML DOM is shown in code listing \ref{lst:htmldom}. The tag will be uploaded onto the distributed ledger so that users are notified of the data being crawled. Moreover, the Consent Tagging Processor will together with other modules mask the content where the consent configuration specified by the user blocks certain crawlers. For circumstances where there are services such as Content Delivery Network (CDN) involved, additional components may be required, for instance Bot Management and Service Workers.

\begin{lstlisting}[language=HTML, basicstyle=\footnotesize, label={lst:htmldom}, caption={Example of consent information injected in HTML DOM elements as attributes.}]
<div class="article-contents">
  <p class="post-body"
    consent-tag-hash="ab4a39e4fc8118cbb37c..."
    consent-tag-sig="646d3d1079b5ac19ea5b...">
    I really love this post.
  </p>
</div>
\end{lstlisting}

Once we scrape the web page, the web crawler will keep the link between the tag and data throughout later processes of cleaning and preprocessing. Once the data is acquired and used by another entity, the Consent Resolver logs the tags of data on a distributed ledger. If a user requests consent withdrawal on the distributed ledger, Consent Resolver should remove the corresponding data of specific tags being requested from the original and also downstream datasets.

\begin{figure*}[htbp]
    \centerline{\includegraphics[width=\textwidth]
    {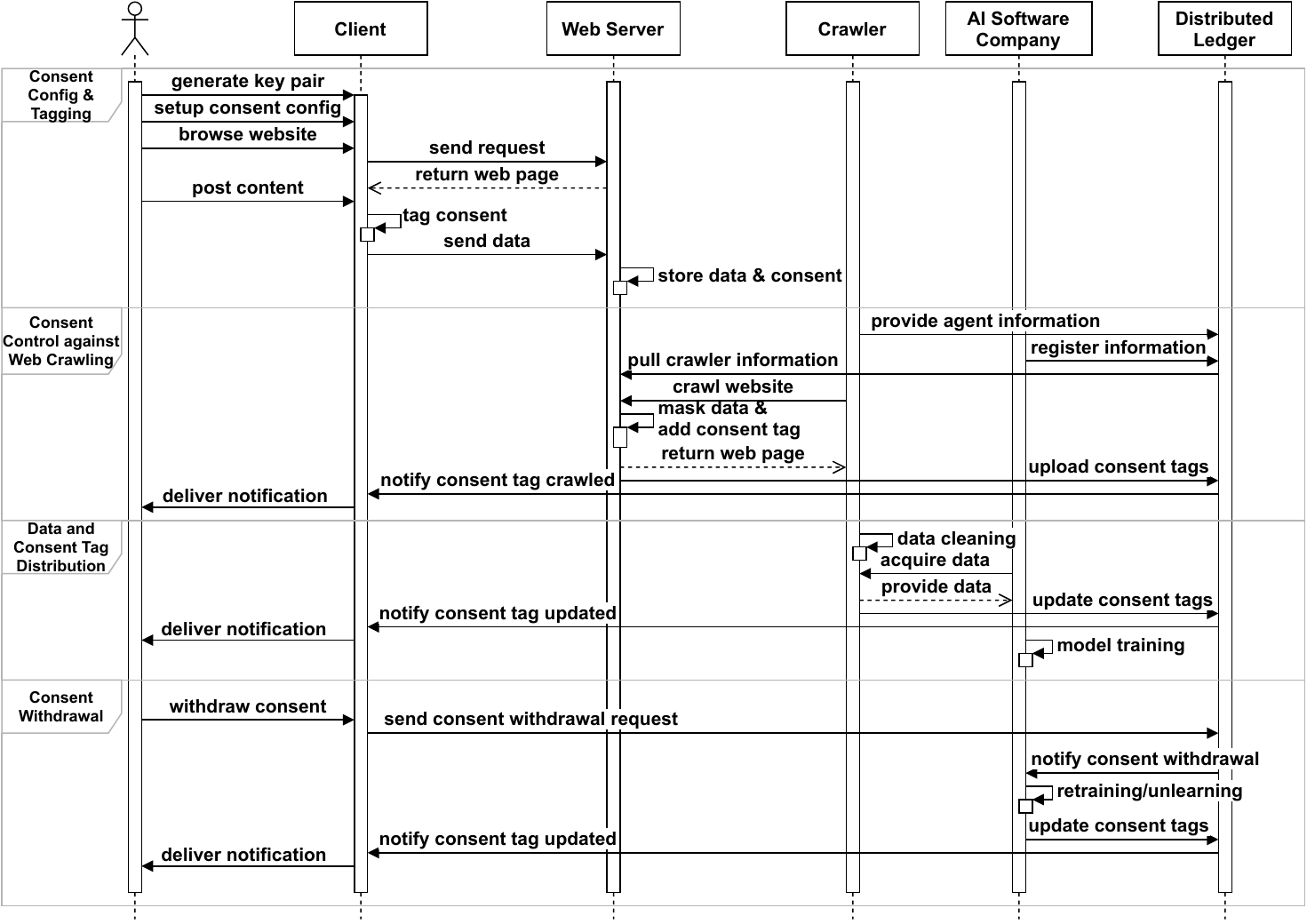}}
    \caption{System Mechanism. The complete lifecycle contains four processes, including Consent Configuration and Tagging, Consent Control against Web Crawling, Data and Consent Tag Distribution, and Consent Withdrawal.}
    \label{process}
\end{figure*}

\noindent \textbf{Machine Learning Side.}
AI software companies acquire datasets from web crawling entities, or they may themselves perform web scraping activities. They then train machine learning models using these datasets. The company keeps the records of consent linked with the data in the Consent Records repository. If the user requests consent withdrawal, the company will remove the corresponding data from the dataset. However, merely removing data from datasets might not be enough, as the data is already trained into the weights of models, which might still invoke legal implications related to privacy or copyrights. Therefore, the Retraining/Unlearning Trigger should be activated by the consent withdrawal requests, and invoke the re-training or unlearning process of machine learning models. The retraining/unlearning process should be logged on the distributed ledger for transparency.

\noindent \textbf{Distributed Ledger Side.}
The distributed ledger maintains three components, including i) Consent Logger, which stores the tags of data being crawled, used for training, and requested consent withdrawal; ii) Agent Configurations, which store configurations of web crawlers such as their user agent information and IP ranges, and web servers can use this information to identify crawlers and enforce relevant policies against crawlers, and; iii) Consent Request Handler, which accepts consent withdrawal requests from users and notifies relevant parties. %
The distributed ledger provides availability, consistency, transparency, and decentralization for consent management, and can prevent the nature of a single point of failure and centralized power of a centralized platform. 

\subsection{System Mechanism}
\label{sec:system-mechanism}

The system mechanism is depicted in figure~\ref{process}, showcasing four processes: Consent Configuration and Tagging, Consent Control against Web Crawling, Data and Consent Tag Distribution, and Consent Withdrawal.

\noindent \textbf{Consent Configuration and Tagging.}
The user should generate at least one key pair for signing the consent tags. Optionally, the user may set up a consent configuration to specify which crawlers are allowed to scrape their online data or skip to allow all crawlers by default. The key pair and consent configuration are stored in the Integrated Consent Tagging Extension of the client application. Users may generate additional key pairs, or modify consent configuration at any time. Please note that when the user is prompted by the client-side interface, such as an HTML input element, to send data, the client-side program, when initiating the request, may specify the request as \textit{non-crawlable} in the HTTP header, which will be read by consent tagging extension, so that no consent configuration and tags will be attached together with the data. This can be used for data such as passwords and other non-public content.

If the field is not marked as non-crawlable, the consent tagging extension will hash and sign data using the private key, and attach the resulting hash and signature with the consent configuration in a header of the HTTP request. The data, consent configuration, hash, signature, signing key pair, and metadata such as URL and time are stored locally. The web server stores the consent configuration, hash, signature, and their link with the data, in the database.

\noindent \textbf{Consent Control against Web Crawling.} 
The Agent Configurations of crawlers can be read by the website owners, and the configurations can be regularly updated by the crawling entities. The security mechanism of distributed ledger ensures the authenticity of these configurations~\cite{xu2019blockchain}. When the server is being crawled by a web crawler, the crawler first goes to \textit{robots.txt}\footnote{\url{https://www.rfc-editor.org/rfc/rfc9309}} to read which portions of the website it is allowed to visit. These instructions are set by the website owner, and the granularity is commonly at the level of directories. Then the crawler will request each URL it wishes to scrape. The request contains the client IP address and the "User-agent" header\footnote{\url{https://developer.mozilla.org/en-US/docs/Web/HTTP/Headers/User-Agent}},
by which the server can identify the crawler. The crawler may also add a signature on the current timestamp to the HTTP headers in order to prove its authenticity. Anti-crawler strategies are enforced to protect the website against malicious crawlers and illegal visits. If certain trusted crawlers e.g. \textit{Googlebot}\footnote{\url{https://developers.google.com/search/docs/crawling-indexing/googlebot}} are allowed by the website, the server should retrieve the data from the database using techniques such as \textit{Object-relational mapping (ORM)}. Corresponding consent configuration and consent tags should be extracted together, and the server should filter these data based on consent configuration. If the requesting crawler is not allowed in certain data, such data should be excluded from the response by techniques such as masking the HTML elements. If the consent configuration does not reject the crawler, the server should inject the consent tags into the response by adding relevant \textit{consent tag} attribute in HTML elements, similar to the example in listing \ref{lst:htmldom}. The server should also log the activity on a distributed ledger to associate the consent tag with the crawler, and the user will receive notifications about activities related to their data.

\noindent \textbf{Data and Consent Tag Distribution.}
After being scraped by the crawler, the data will go through a number of steps before being included in datasets, and consent tags should be always kept with the data throughout the process. Aggregated data should be kept with a collection of consent tags of the original data. Every time the data is transferred from one party to another, or being trained into a machine learning model, the transfer or training information should be logged to the entries of the data on the distributed ledger, and the user will be notified accordingly. Furthermore, large-volume downstream distribution should be also logged and tracked on the distributed ledger. This traceability can be further augmented using watermarking technology\footnote{\url{https://huggingface.co/blog/alicia-truepic/identify-ai-generated-content}}.

\noindent \textbf{Consent Withdrawal.}
At any time, users are able to request withdrawal of their consent through a distributed ledger, without revealing their true identity. Users can view the journey of their data using the client-side Consent Management Endpoint, which connects to the distributed ledger to query the information based on local records. If users intend to withdraw consent to certain data, they can send the signing public key, the hash of the data, and a signature of a challenge for withdrawal to the distributed ledger. Once all this information is automatically verified to be authentic by the Consent Request Handler on the distributed ledger, all parties holding the original or downstream data will receive the notification about this consent withdrawal, and once their removal of data is completed, they should report the completion to the Consent Logger, which will subsequently notify the user. The authority and the public can audit the relevant parties based on the consent records throughout the process.

\section{Implementation}
\label{sec:implementation}

We implement a proof-of-concept system to validate the feasibility of architectural design. In this section, we describe the main software libraries and core operations of modules within each layer.

\subsection{Web Side}
The base system of a web server is implemented as a streamlined social media website based on LoopBack 4\footnote{\url{https://loopback.io/}}. The website is deployed onto a Google Cloud Platform 4 vCPU 16GB memory virtual machine with a Debian 12 operating system. PostgreSQL\footnote{\url{https://www.postgresql.org/}} v15.0 is used as database, in combination with Redis\footnote{\url{https://redis.io/}} v7.2 for the REST-level caching. On top of the base system, the Consent Tagging Processor is facilitated with an additional database table, coupled with injector mechanisms in the HTTP handlers of the base system. Algorithm \ref{consent-tagging-processor-user-request} demonstrates the operations of the Consent Tagging Processor when a user request containing user data is received, effectively linking the consent information with data. Algorithm \ref{consent-tagging-processor-crawler-request} demonstrates the operations of masking data and injecting consent information into response when the web page is being crawled by crawlers.

We employ Scrapy\footnote{\url{https://github.com/scrapy/scrapy}} v2.11.0 as the crawler in combination with Splash\footnote{\url{https://github.com/scrapinghub/splash}} v3.5.0 to enable JavaScript rendering capability. The crawler extracts elements from web pages and outputs data in gzip format.

\begin{algorithm}
\caption{\small Consent Tagging Processor on receiving user POST request}\label{consent-tagging-processor-user-request}
\begin{algorithmic}
\footnotesize
\State data = request.data
\State dataId = ORM.save(data)
\If{crawlable == True}
    \State consentId = consentStore.save(dataId, request.consentInfo)
    \State ORM.updateConsentId(dataId, consentId)
\EndIf
\end{algorithmic}
\end{algorithm}

\begin{algorithm}
\caption{\small Consent Tagging Processor on receiving crawler request}\label{consent-tagging-processor-crawler-request}
\begin{algorithmic}
\footnotesize
\State crawlerInfo = CrawlerInfo.get(request.headers["user-agent"])
\State data = ORM.extract(dataIds)
\State consentInfo = consentStore.extract(dataIds)
\For{d in data}
\If{!checkConsent(consentInfo[d.id], crawlerInfo)}
\State d = mask(d)
\Else
\State d = addConsentInfo(d, consentInfo[d.id])
\EndIf
\EndFor
\State response.send(data)
\end{algorithmic}
\end{algorithm}

\subsection{Client Side}

The frontend is implemented %
using React\footnote{\url{https://react.dev/}} 16.8.6. The Integrated Consent Tagging Extension is implemented as a part of a Chrome browser extension\footnote{\url{https://developer.chrome.com/docs/extensions/}} with manifest v2. The Consent Management Endpoint is housed within the same extension, using web3.js\footnote{\url{https://github.com/web3/web3.js}} v1.9.0. As shown in Algorithm \ref{consent-tagging-extension-user-request}, the extension signs the data on behalf of the user when a request is sent, and stores relevant information for tracking and withdrawal of consent. The extension uses stored information to query the distributed ledger for consent withdrawal. The hash algorithm used is Keccak-256\footnote{\url{https://docs.web3js.org/api/web3-utils/function/sha3/}}, and the digital signature scheme is Elliptic Curve Digital Signature Algorithm (ECDSA)\footnote{\url{https://web3js.readthedocs.io/en/v1.9.0/web3-eth.html\#sign}} with named curve NIST P-384\footnote{\url{https://csrc.nist.gov/pubs/fips/186-4/final}}.

\begin{algorithm}
\caption{\small Integrated Consent Tagging Extension on sending request}\label{consent-tagging-extension-user-request}
\begin{algorithmic}
\footnotesize
\Function{webRequest.onBeforeSendHeaders.listener}{}
\If{request.method == "POST" and "non-crawlable" not in request.headers}
    \State h = hash(request.data)
    \State s = sign(privKey, h)
    \State request.headers.add({h,s,consentConfig})
    \State localStorage.add({h,s,pubKey,consentConfig,metadata})
\EndIf
\EndFunction
\end{algorithmic}
\end{algorithm}

\subsection{Machine Learning Side}
As the focus of the architecture is on the tracking and withdrawal of user content instead of training of machine learning models, the functional modules of the Machine Learning side are not implemented into a real system. We implemented a retraining trigger using bash, and a Consent Records module bridged with the distributed ledger by the endpoint implemented using web3.js. The endpoint will receive emitted consent withdrawal message from the distributed ledger, and deliver the message to the Trigger.

\subsection{Distributed Ledger}

We selected the Ethereum blockchain as the distributed ledger platform and implemented the modules as smart contracts using Solidity 0.8.20. The Consent Logger smart contract keeps a mapping of all consent tags with their hashes, signatures, custodians, and associated requests. The Consent Request Handler serves as an interface for users to engage with the Logger for consent management. Additionally, the Agent Configurations smart contract maintains the configurations for Crawlers. Web Servers can query this contract to obtain crawler configurations and their future updates, subsequently enforcing relevant server-side policies.

\section{Evaluation}
\label{sec:evaluation}

The implemented proof-of-concept system runs smoothly throughout the lifecycle of user data. The consent information is correctly embedded into request headers, subsequently received by the web server, and scraped by the crawler. This information is then incorporated into the AI model's training data. Users are promptly notified about their data usage via a distributed ledger, enabling them to request data withdrawal using cryptographical mechanisms without the necessity of authenticating their identity.

The framework is compatible with the existing internet infrastructure. In our assessments, it effectively accommodates requests of types POST, PUT, and PATCH.
According to our evaluation results, the proof-of-concept implementation does not inherently cater to WebSocket requests, but this capability can be realized by intercepting WebSocket connections prior to their establishment. This also holds for custom methods.

Due to the extensibility of HTTP and HTML, our framework, which is built upon this spirit, also possesses considerable extensibility. The injected consent headers support expanded fields evolved conventions, and different cryptographic data. Likewise, the injected consent attributes of DOM elements are extensible.

The framework seamlessly integrates with the current internet infrastructure. Components, such as the client-side Consent Tagging Extension and the web-side Consent Tagging Processor, all function correctly with websites or clients that have not adopted the framework.

\begin{figure}[htbp]
    \centerline{\includegraphics[width=0.7\columnwidth]
    {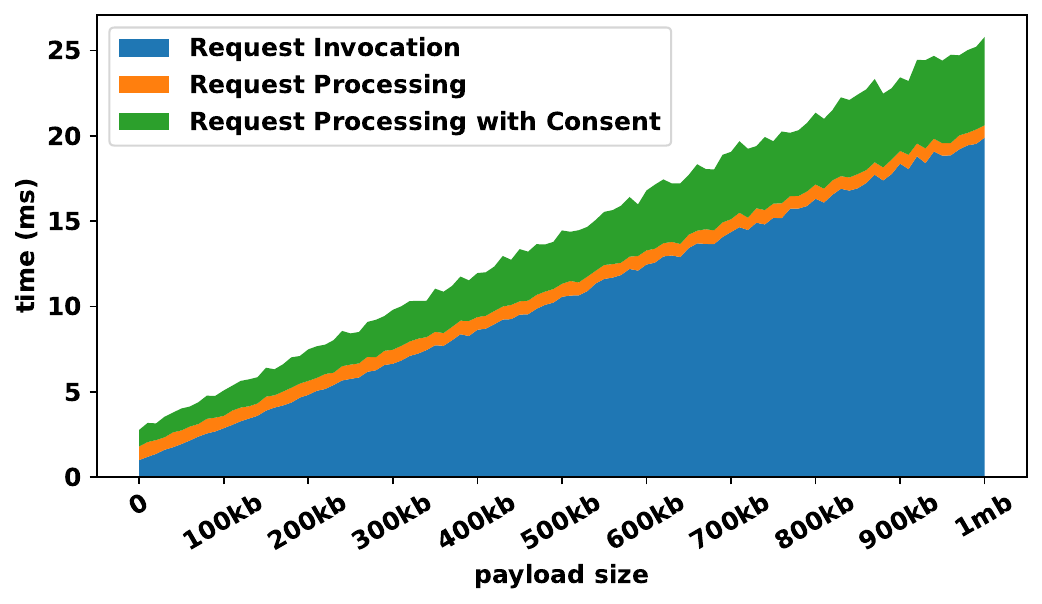}}
    \caption{Time cost for Request Invocation, Request Processing, and Request Processing with Consent, across varying payload sizes. The time associated with Request Processing with Consent indicates the additional client-side overhead introduced by the framework.}
    \label{client-perf}
\end{figure}

Furthermore, to assess the impact of the framework on the existing applications, we ran a series of micro-benchmarking on the implemented system. The client-side overhead is shown in Figure \ref{client-perf}. Each data point depicted in the figure represents the mean value derived from 20 individual runs to ensure statistical reliability. The framework causes additional overhead for sending requests from the client side. However, the overhead for a payload size of 1mb is only around 5ms, which we believe is negligible, particularly compared with the non-avoidable overhead of around 20ms from the request invocation from frontend code.

\begin{figure}[htbp]
    \centerline{\includegraphics[width=0.7\columnwidth]
    {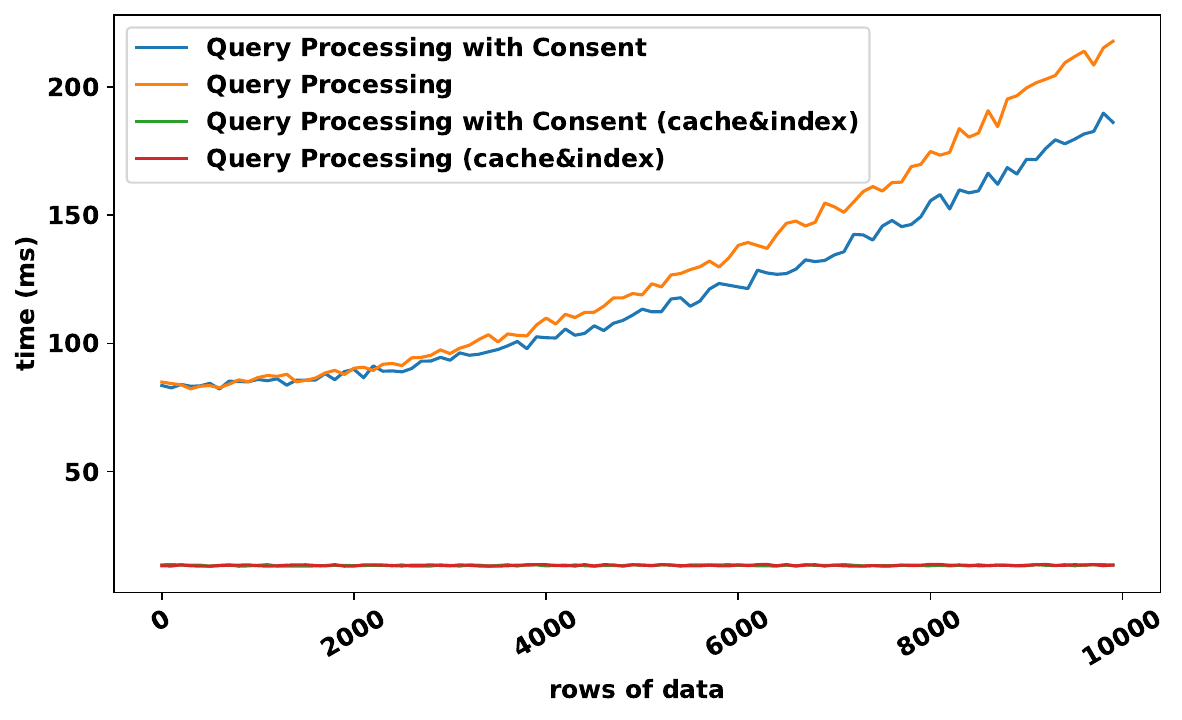}}
    \caption{Time cost for Query Processing, across varying number of data rows. Time associated with Query Processing with Consent indicates the server-side overhead when using the framework.}
    \label{server-perf}
\end{figure}

For backend overhead, the results are shown in Figure \ref{server-perf}. As modern web services adopt a range of technologies such as caching and indexing, to optimize the performance of queries, we initially disabled these features to compare the difference in time cost. Though the framework added time cost when querying a relatively large amount of data, the increment was merely around 20ms, which could be deemed negligible. Moreover, when the optimization features are enabled, the time costs are nearly identical, demonstrating that our framework is unlikely to introduce additional burdens on typical web services.

In addition, our evaluation of the framework extends to the transaction capacity on the selected distributed ledger, Ethereum. Our findings indicate that a single transaction, adhering to a typical gas limit of 30M, can encapsulate over 47,000 consent information entries. This scenario operates under the assumption that web servers aggregate and upload consent information in bulk within a singular transaction, executed after specified intervals. Our testing, conducted on the Goerli testnet, received confirmation on the network within an average of 23 seconds. In an optimal scenario utilizing consortium blockchains, which have considerably higher transactions per second (TPS) rates, the bottleneck observed on Ethereum will become less constraining for this framework.

\section{Discussion}
\label{sec:discussion}

This work introduces a framework dedicated to strengthening the privacy and copyright protections of online users, overcoming the limitations inherent in existing approaches predicated on corporate choices. Rather than endeavoring to create an entirely new internet architecture, the objective of this framework is to augment the existing internet infrastructure. This is achieved by harnessing its inherent extensibility to enable enhanced privacy and copyright capabilities. Moreover, the extensions to HTTP and HTML are engineered to be perceptible only to client, server, and crawler applications, while remaining invisible to users, which facilitates a seamless and non-disruptive upgrade. The decentralized architecture guarantees the availability of the framework, ensuring users retain continual access to consent management features, independent of the availability of specific service providers or authorities.

Additionally, this framework introduces consent tags which is based on cryptographic methods, ensuring that users are not required to provide additional personal information, thereby avoiding a compromise of privacy while exercising their rights.

Nonetheless, this framework is designed with good actors in mind. Similar to voluntary measures like providing options for website administrators to restrict crawlers, or adhering to the Robots Exclusion Protocol, the efficacy of this framework hinges on the willingness of companies or legislations to adopt it. Moreover, executing data deletion post-withdrawal requests might necessitate auditing by authoritative third parties. We acknowledge this as a limitation of our framework. Yet, given the recent surge in public concerns regarding privacy and copyright in the context of GenAI, coupled with the demonstrated willingness for self-regulation by various companies, we posit that this framework constitutes a significant stride towards building a responsible web.

\section{Related Work}
\label{sec:related-work}

The challenges of user data control in the digital realm have been addressed from various angles in the literature.

Considering the diversity of web applications and the ``data silos'' they have created, SOLID~\cite{sambra2016solid} aims to address data ownership and privacy concerns by offering data stores under user control. The decentralized \textit{Pods}, allow users to manage data access by applications or individuals, with the option to revoke permissions. Although SOLID lays the groundwork for user-centric data management, its adoption is still in the early stages with challenges in broader web integration and application support, indicating a need for more integrative solutions within the existing web ecosystem.

Chen et al.~\cite{Chen2015Self} utilized the XACML policy language to craft user-customized access control policies. This methodology offers a robust mechanism for data sharing in a secure environment. Whilst, the current Web-based applications usually do not support such schemes to prevent data crawling, hence, we present a multi-layered solution to address this challenge.

The decentralized nature of blockchain technology offers promising avenues for secure data sharing. Several researchers have delved into harnessing blockchain for ensuring data privacy and integrity and fostering user control~\cite{8292361, nguyen2023blockchainempowered, 10.1145/3350546.3352561}. Specifically, blockchain has been leveraged for copyright management~\cite{8377886, SAVELYEV2018550}, which can be generalized to stakeholders' consent management over their data. 

\section{Conclusion}
\label{sec:conclusion}

The World Wide Web (WWW), serving as a vital source of information for individuals, has posed challenges in safeguarding user privacy and copyrights, especially amidst the recent rise of Generative AI. The prevailing WWW falls short of providing adequate mechanisms for consent withdrawal or data copyright claims, leaving both users and data holders at a disadvantage. This paper presents a user-controlled consent tagging framework, built on top of the existing internet infrastructure, leveraging the extensibility of HTTP and HTML, alongside distributed ledger technology, to address these concerns. Users can tag their online data at the point of transmission, monitor its usage, and request consent withdrawal from data holders. Through the evaluation of a proof-of-concept system, the effectiveness of the framework is substantiated. We believe that this work paves the way for a more responsible, user-centric web ecosystem.



\begin{thebibliography}{10}
\providecommand{\url}[1]{#1}
\csname url@samestyle\endcsname
\providecommand{\newblock}{\relax}
\providecommand{\bibinfo}[2]{#2}
\providecommand{\BIBentrySTDinterwordspacing}{\spaceskip=0pt\relax}
\providecommand{\BIBentryALTinterwordstretchfactor}{4}
\providecommand{\BIBentryALTinterwordspacing}{\spaceskip=\fontdimen2\font plus
\BIBentryALTinterwordstretchfactor\fontdimen3\font minus \fontdimen4\font\relax}
\providecommand{\BIBforeignlanguage}[2]{{%
\expandafter\ifx\csname l@#1\endcsname\relax
\typeout{** WARNING: IEEEtran.bst: No hyphenation pattern has been}%
\typeout{** loaded for the language `#1'. Using the pattern for}%
\typeout{** the default language instead.}%
\else
\language=\csname l@#1\endcsname
\fi
#2}}
\providecommand{\BIBdecl}{\relax}
\BIBdecl

\bibitem{berners1999weaving}
T.~Berners-Lee, \emph{Weaving the Web: The original design and ultimate destiny of the World Wide Web by its inventor}.\hskip 1em plus 0.5em minus 0.4em\relax Harper San Francisco, 1999.

\bibitem{khan2022subjects}
M.~Khan and A.~Hanna, ``The subjects and stages of ai dataset development: A framework for dataset accountability,'' 2022.

\bibitem{hacker2023regulating}
P.~Hacker, A.~Engel, and M.~Mauer, ``Regulating chatgpt and other large generative ai models,'' in \emph{Proceedings of the 2023 ACM Conference on Fairness, Accountability, and Transparency}, 2023, pp. 1112--1123.

\bibitem{gdpr}
\BIBentryALTinterwordspacing
``Regulation (eu) 2016/679 of the european parliament and of the council of 27 april 2016 on the protection of natural persons with regard to the processing of personal data and on the free movement of such data, and repealing directive 95/46/ec (general data protection regulation),'' 2016. [Online]. Available: \url{https://eur-lex.europa.eu/legal-content/EN/TXT/?uri=CELEX%3A02016R0679-20160504}
\BIBentrySTDinterwordspacing

\bibitem{dikaiakos2005investigation}
M.~D. Dikaiakos, A.~Stassopoulou, and L.~Papageorgiou, ``An investigation of web crawler behavior: characterization and metrics,'' \emph{Computer Communications}, vol.~28, no.~8, pp. 880--897, 2005.

\bibitem{manning2008introduction}
C.~D. Manning, P.~Raghavan, and H.~Sch{\"u}tze, ``Introduction to information retrieval,'' 2008.

\bibitem{brickell2008cost}
J.~Brickell and V.~Shmatikov, ``The cost of privacy: destruction of data-mining utility in anonymized data publishing,'' in \emph{Proceedings of the 14th ACM SIGKDD international conference on Knowledge discovery and data mining}, 2008, pp. 70--78.

\bibitem{manku2007detecting}
G.~S. Manku, A.~Jain, and A.~Das~Sarma, ``Detecting near-duplicates for web crawling,'' in \emph{Proceedings of the 16th international conference on World Wide Web}, 2007, pp. 141--150.

\bibitem{garcia2015data}
S.~Garc{\'\i}a, J.~Luengo, and F.~Herrera, \emph{Data preprocessing in data mining}.\hskip 1em plus 0.5em minus 0.4em\relax Springer, 2015, vol.~72.

\bibitem{roh2019survey}
Y.~Roh, G.~Heo, and S.~E. Whang, ``A survey on data collection for machine learning: a big data-ai integration perspective,'' \emph{IEEE Transactions on Knowledge and Data Engineering}, vol.~33, no.~4, pp. 1328--1347, 2019.

\bibitem{7163042}
Y.~Cao and J.~Yang, ``Towards making systems forget with machine unlearning,'' in \emph{2015 IEEE Symposium on Security and Privacy}, 2015, pp. 463--480.

\bibitem{5484614}
M.~Karasuyama and I.~Takeuchi, ``Multiple incremental decremental learning of support vector machines,'' \emph{IEEE Transactions on Neural Networks}, vol.~21, no.~7, pp. 1048--1059, 2010.

\bibitem{jaderberg2014synthetic}
M.~Jaderberg, K.~Simonyan, A.~Vedaldi, and A.~Zisserman, ``Synthetic data and artificial neural networks for natural scene text recognition,'' \emph{arXiv preprint arXiv:1406.2227}, 2014.

\bibitem{carlini2021extracting}
N.~Carlini, F.~Tramer, E.~Wallace, M.~Jagielski, A.~Herbert-Voss, K.~Lee, A.~Roberts, T.~Brown, D.~Song, U.~Erlingsson \emph{et~al.}, ``Extracting training data from large language models,'' in \emph{30th USENIX Security Symposium (USENIX Security 21)}, 2021, pp. 2633--2650.

\bibitem{shokri2017membership}
R.~Shokri, M.~Stronati, C.~Song, and V.~Shmatikov, ``Membership inference attacks against machine learning models,'' in \emph{2017 IEEE symposium on security and privacy (SP)}.\hskip 1em plus 0.5em minus 0.4em\relax IEEE, 2017, pp. 3--18.

\bibitem{wachter2019data}
S.~Wachter, ``Data protection in the age of big data,'' \emph{Nature Electronics}, vol.~2, no.~1, pp. 6--7, 2019.

\bibitem{zhang2023right}
D.~Zhang, P.~Finckenberg-Broman, T.~Hoang, S.~Pan, Z.~Xing, M.~Staples, and X.~Xu, ``Right to be forgotten in the era of large language models: Implications, challenges, and solutions,'' \emph{arXiv preprint arXiv:2307.03941}, 2023.

\bibitem{renieris2023beyond}
E.~M. Renieris, \emph{Beyond data: reclaiming human rights at the dawn of the metaverse}.\hskip 1em plus 0.5em minus 0.4em\relax MIT Press, 2023.

\bibitem{xu2019blockchain}
X.~Xu, I.~Weber, and M.~Staples, ``Blockchain in software architecture,'' \emph{Architecture for Blockchain Applications}, pp. 83--92, 2019.

\bibitem{sambra2016solid}
A.~V. Sambra, E.~Mansour, S.~Hawke, M.~Zereba, N.~Greco, A.~Ghanem, D.~Zagidulin, A.~Aboulnaga, and T.~Berners-Lee, ``Solid: a platform for decentralized social applications based on linked data,'' \emph{MIT CSAIL \& Qatar Computing Research Institute, Tech. Rep.}, 2016.

\bibitem{Chen2015Self}
S.~Chen, D.~Thilakanathan, D.~Xu, S.~Nepal, and R.~Calvo, ``Self protecting data sharing using generic policies,'' in \emph{2015 15th IEEE/ACM International Symposium on Cluster, Cloud and Grid Computing}, 2015, pp. 1197--1200.

\bibitem{8292361}
X.~Liang, J.~Zhao, S.~Shetty, J.~Liu, and D.~Li, ``Integrating blockchain for data sharing and collaboration in mobile healthcare applications,'' in \emph{2017 IEEE 28th Annual International Symposium on Personal, Indoor, and Mobile Radio Communications (PIMRC)}, 2017, pp. 1--5.

\bibitem{nguyen2023blockchainempowered}
L.~T. Nguyen, L.~D. Nguyen, T.~Hoang, D.~Bandara, Q.~Wang, Q.~Lu, X.~Xu, L.~Zhu, P.~Popovski, and S.~Chen, ``Blockchain-empowered trustworthy data sharing: Fundamentals, applications, and challenges,'' 2023.

\bibitem{10.1145/3350546.3352561}
\BIBentryALTinterwordspacing
S.~Rouhani and R.~Deters, ``Blockchain based access control systems: State of the art and challenges,'' in \emph{IEEE/WIC/ACM International Conference on Web Intelligence}, ser. WI '19.\hskip 1em plus 0.5em minus 0.4em\relax New York, NY, USA: Association for Computing Machinery, 2019, p. 423–428. [Online]. Available: \url{https://doi.org/10.1145/3350546.3352561}
\BIBentrySTDinterwordspacing

\bibitem{8377886}
Z.~Meng, T.~Morizumi, S.~Miyata, and H.~Kinoshita, ``Design scheme of copyright management system based on digital watermarking and blockchain,'' in \emph{2018 IEEE 42nd Annual Computer Software and Applications Conference (COMPSAC)}, vol.~02, 2018, pp. 359--364.

\bibitem{SAVELYEV2018550}
\BIBentryALTinterwordspacing
A.~Savelyev, ``Copyright in the blockchain era: Promises and challenges,'' \emph{Computer Law \& Security Review}, vol.~34, no.~3, pp. 550--561, 2018. [Online]. Available: \url{https://www.sciencedirect.com/science/article/pii/S0267364917303783}
\BIBentrySTDinterwordspacing

\end{thebibliography}
\end{document}